\documentclass[journal]{IEEEtran}

\usepackage{verbatim}
\usepackage{graphicx}
\usepackage[cmex10]{amsmath}
\usepackage{amssymb,amsfonts}
\usepackage{multirow}
\usepackage{theorem}

\newcommand\MYhyperrefoptions
{
	bookmarks=false, 
	bookmarksnumbered=true,
	pdfpagemode={UseOutlines},
	plainpages=false,
	pdfpagelabels=true,
	colorlinks=true,
	linkcolor={black},
	citecolor={black},
	urlcolor={black},
	pdftitle={}, 
	pdfcreator={LaTex},
}

\usepackage{cite}
\usepackage[\MYhyperrefoptions,breaklinks=true,]{hyperref}
\usepackage{breakurl}
\usepackage[tight,footnotesize]{subfigure}
\usepackage{algorithm}
\usepackage{algorithmic}

\usepackage[justification=centering]{caption}  
\usepackage{bbm}

\usepackage{color}

\begin{document}

\title{
{
Cooperative MARL-Based Energy-Efficient Power Control for Two-Hop Relaying Networks
}
}

\author{
    Xinyu~Qu, 
    Yuanzhe~Geng, 
    Erwu~Liu, 
    and Rui~Wang 
    
    \thanks{
    Xinyu Qu, Yuanzhe~Geng, Erwu Liu, and Rui Wang are with the College of Electronics and Information Engineering, Tongji University, Shanghai 201804, China, E-mails: xinyuqu@tongji.edu.cn, yuanzhegeng@tongji.edu.cn, erwu.liu@ieee.org, ruiwang@tongji.edu.cn
    .
    }
    
    \thanks{
    Xinyu Qu and Yuanzhe~Geng are co-first authors. Erwu Liu is the corresponding author.
        
    }

}


\maketitle

\begin{abstract}
In this paper, we study a cooperative game in the cooperative communication network, where each relay makes decisions autonomously and aims to achieve the same optimization objective of maximizing energy efficiency.  
We consider the non-ideal situation where instantaneous channel state information (CSI) is difficult to obtain and only partially observable outdated CSI is available. 
To solve this game problem, we define a delayed reward-based state-action value function and propose a multi-agent deep Q network learning framework. 
Then, we prove analytically that utilities obtained by game-theoretic approaches with the instantaneous CSI serve as upper bounds for those of the proposed method. 
Simulation results reveal that our approach considerably outperforms its potential alternatives and is only about 5.2\% away from the optimal solution.

\end{abstract}


\section{Introduction}\label{sect_intro}
Cooperative diversity has been a promising technique to improve the transmission of information in recent years \cite{10857950}, along with the increasing concentration on the energy consumption of the system.
To improve the energy efficiency of cooperative communication systems, the optimization of power allocation is one of the most straightforward ways. 
Currently, traditional methods usually make artificial assumptions on the underlying channels, and then employ mathematical tools, such as convex optimization, to derive the optimal solution that maximizes the energy efficiency \cite{10153599, 9675818}. 
%
%
    These efforts assume an implicit central controller that makes schedules for the whole system with global information, which consequently limits their application in distributed systems.

To this end, multi-agent reinforcement learning (MARL) is proposed to solve cooperative tasks where each node in the communication system acts as an independent player/agent. 
    For example, Lv \textit{et.al.}~\cite{9645220} maximized the energy  efficiency of the secure short-packet transmission in massive machine-type communications, and proposed a decentralized stateless Q-learning method.
    Galkin \textit{et.al.}~\cite{9838809} studied the optimization of unmanned aerial vehicles' flight trajectories, and employed MARL to maximized the total energy efficiency of the system. 
    These studies showed the potential of MARL for solving cooperative tasks in the field of communications.   
    In practical applications, another challenge is that the instantaneous channel state information (CSI) may be difficult to obtain due to the rapid change of channel states.  
However, few existing MARL-based cooperative works consider such 
non-ideal situation. 
The design of RL systems and how RL performs under such non-ideal conditions require further research. 

Motivated by this, we study a cooperative game between relays in a two-hop cooperative network in this paper. 
With partially observed outdated CSI, each relay acts as an independent agent and cooperates to maximize the system's energy efficiency by determining its own transmit power. 
To solve this cooperative game problem, we define a delayed reward-based state-action value function and propose a multi-agent deep Q network (DQN) learning framework. 
We show that the utilities of the game-theoretic approach with instantaneous CSI serve as upper bounds for those of the RL-based approach with outdated CSI, for both long-period and one-step decisions in communication environments. 
Simulation results demonstrate that the proposed distributed scheme outperforms existing learning and non-learning benchmarks in non-ideal environments, with a marginal deviation of approximately 5.2\% from the game-theoretic optimum in ideal situations.

\section{System Model and Game Problem}\label{sect model}
In this section, we first introduce the model of our two-hop cooperative communication network. 
Then, we establish the cooperative game for maximizing energy efficiency. 

\subsection{Communication Model}
Consider a typical two-hop wireless relay network, where there is a source $S$, a destination $D$, and a group of $K$ relays. 
The source communicates with the destination with the assistance of relays, where the amplify-and-forward (AF) protocol is adopted.


The relays use the half-duplex signaling mode. 
Therefore, communication from the source $S$ to the destination $D$ via relay $R_k, 1 \leq k \leq K$, takes two time slots.
In the first time slot, the source broadcasts its signal.
Assuming that the direct link is obstructed between the source and destination.
Only relays receive this transmission, and the received signal at each relay $R_k$ is written as 
\begin{equation} 
    y_{sk}(t) = \sqrt{P_s} h_{sk}(t) x(t) + n_k(t), 
\end{equation}
where $P_s$ represents the transmit power of the source; 
$x(t)$ represents data symbol; 
$h_{sk}(t)$ represents the channel gain between the source and relay $R_k$, which is a complex Gaussian random variable with zero mean and variance $\sigma_{sk}^2$; 
and $n_k(t)$ is the complex Gaussian noise at relay $R_k$, and it has zero mean and variance $\sigma_k^2$.

Further, to characterize the temporal correlation between consecutive time slots in each channel, we employ the Gaussian Markov block fading auto-regressive model \cite{5710995, 9568962}:  
\begin{equation} \label{channel state} 
    h_{ij}(t)=\rho h_{ij}(t-1)+\sqrt{1-\rho^2} \zeta(t),  
\end{equation}
where the subscript $_{i, j}$ represents two different nodes, $\rho$ is the normalized channel correlation coefficient, and $\zeta(t)\sim\mathcal{CN}(0,\sigma_j^2)$ represents the error variable which is uncorrelated with $h_{ij}(t)$.

In the second time slot, the source is silent, and each relay $R_k$ amplifies and forwards its detected signal to the destination.
Then, the received signal at the destination can be written as
\begin{equation}
    y_{d}(t)=\sum_{k=1}^{K}\sqrt{P_k} h_{kd}(t)x_{sk}(t) + n_d(t)
\end{equation}
where $P_k$ is the transmit power of the relay $R_k$, and $x_{sk}(t)=y_{sk}(t)/\|y_{sk}(t)\|$. 
Similarly, $h_{kd}(t)$ represents the channel gain between relay $R_k$ and the destination, which is also a complex Gaussian random variable with zero mean and variance $\sigma_{kd}^2$. 
$n_d(t)$ is the complex Gaussian noise at the destination with zero mean and variance $\sigma_d^2$.
Then, the received signal above can be further written as 
\begin{equation}\begin{aligned}
y_{d}(t) = 
&\sum_{k=1}^{K} \sqrt{P_k} h_{kd}(t) \alpha_k \sqrt{P_s} h_{sk}(t) x(t) \\
&+ \sum_{k=1}^{K} \sqrt{P_k} h_{kd}(t) \alpha_k n_k(t) + n_d(t)
\end{aligned}\end{equation}
with $\alpha_k=(P_s \|h_{sk}(t)\|^2 + \sigma_k^2)^{-\frac{1}{2}}$ being the amplification factor for conciseness.

The output SNR at the destination can be written as \cite{6954557}
\begin{equation}
\gamma = \frac
{\big( \sum\nolimits_{k=1}^{K} \sqrt{P_k} h_{kd} \alpha_k h_{sk} \big)^2 P_s}
{\sum\nolimits_{k=1}^{K} P_k \|h_{kd}\|^2 \alpha_k^2 \sigma_k^2 + \sigma_d^2}.  
\end{equation}
Accordingly, the instantaneous end-to-end channel capacity is:  
\begin{equation}
\begin{aligned}
I = \frac{1}{2}\log_2(1 + \gamma)
=\frac{1}{2}\log_2(1 + \frac
{\big( \sum\nolimits_{k=1}^{K} \sqrt{P_k} h_{kd} \alpha_k h_{sk} \big)^2 P_s}
{\sum\nolimits_{k=1}^{K} P_k \|h_{kd}\|^2 \alpha_k^2 \sigma_k^2 + \sigma_d^2}).
\end{aligned}
\end{equation}

\subsection{Cooperative Game Problem}\label{subsect game model}
Based on the determined transmit power of the source $P_s$, the energy efficiency is defined as follows, which is the ratio of end-to-end channel capacity to the total transmit power consumed by all nodes in the network. 
\begin{equation} \label{energy efficiency}
    EE(P_1, P_2, \dots, P_K) = \frac{I}{P_s + \sum\nolimits_{k=1}^{K} P_k}
\end{equation}

In this work, we investigate a distributed cooperation scenario, where all relays work independently and optimize their own transmit power $P_k$ to help improve the energy efficiency of the system. 
We model the interaction between relays as a cooperative game $\mathcal{G}$, which includes the following parts.  
\begin{itemize}
\item 
\textbf{Player: } The players in this game are from the set of all relay nodes, that is, $R_k, 1 \leq k \leq K$. 
The players take actions simultaneously and aim to achieve higher utilities.

\item
\textbf{Utility: } Considering the characteristic of cooperation, the players are set to have the same utility function, which is the evaluation of energy efficiency as represented in (\ref{energy efficiency}).

\item
\textbf{Policy: } The policy space of each player is defined as a discrete set of transmit power $\{0, 1, 2, \dots, L\} \cdot \frac{P_{k, max}}{L}$, where $P_{k, max}$ represents the maximum transmit power of relay $R_k$, and $L+1$ is the total amount of power levels.
\end{itemize}
%
There are $K$ players in this game, and each player has $L+1$ available policies. 
Obviously, $\mathcal{G}$ is a finite static game, for which there exists the following theorem.

\textit{\textbf{Theorem 1} (Nash \cite{Nash1950}): 
In every finite static game, there exists at least one mixed-strategy Nash equilibrium. 
}

Detailed proofs can be found in \cite{Osborne1997game}. 
$\hfill \square$

This seminal theorem proposed by Nash indicates the existence of the equilibrium state for the game $\mathcal{G}$. 
That is, there is at least one optimal game-theoretic joint action policy $\pi^*_\mathcal{G}$, which provides the joint action $\textbf{P}^*=\{P_1^*, \dots, P_K^* \}$ leading to the maximum energy efficiency $EE^*(\textbf{P}^*)$ at each stage.
However, to achieve the optimal game-theoretic solution, the environment is required to be deterministic, i.e., the global instantaneous CSI should be available.

Unfortunately, in practical situations where channel states change rapidly, it is very possible that the channel state of the current time slot is not obtained immediately, and delay often occurs. 
In such non-ideal environments, each player can only perform actions based on partially outdated CSI. 
The existence of Nash equilibrium can not be guaranteed, and consequently, game-theoretic approaches are not applicable.

\section{MARL-Based Solution}\label{sect solution}
To solve game $\mathcal{G}$ in non-ideal environments, we turn to the RL technique for solutions. 
We take each game player as an independent agent, which makes decisions on power settings according to its partial observation of CSI of the previous time slot. 
Then, each agent receives a reward from the environment. 
We model this process as a Markov decision process (MDP) in this section, and describe our proposed MARL method.

\subsection{Markov Decision Process and System Variables}
RL methods perform policy learning in a trial-and-error manner based on interaction with the external environment. 
The interaction process is established as an MDP, which consists of environment $\mathcal{E}$, state space $\mathcal{S}$, action space $\mathcal{A}$, reward space $\mathcal{R}$, and transition probability $p$. 
At each time slot $t$, the RL agent observes the current state $s_t \in \mathcal{S}_t$ and takes action $a_t \in \mathcal{A}_t$ according to its policy $\pi$.
Afterward, it receives a scalar reward $r_t \in \mathcal{R}_t$ from the environment $\mathcal{E}$, and observes the next state $s_{t+1}$ that follows the transition probability $p(s_{t+1} | s_t, a_t)$. 
The agent's goal is to maximize its cumulative reward $R_t = \sum\nolimits_{i=t}^T\gamma^{i-t}r_i$ that starts from any state $s_t$, where variable $T$ denotes the total number of steps, and $\gamma\in[0,1]$ denotes the discount factor.

Based on the communication and game models in Sect-\ref{sect model}, we design the RL system for the investigated multi-agent environment. 
For each agent, the system variables are designed as follows. 

\begin{itemize}
\item \textbf{State: } 
Without information exchange, the agent only has the outdated channel states associated with it, i.e., a partial observation of the entire environment.
On the other hand, relays act after the source, thus the transmit power of the source can be known to relay agents. 
Thus, we have: 
\begin{equation} 
    \mathcal{S}_t^k \triangleq [h_{sk}(t-1), h_{kd}(t-1), P_s].
\end{equation}

\item \textbf{Action: } 
Each agent needs to decide its transmit power from a discrete set, where the available power is divided into $L$ power levels. 
Since the agent can also choose not to participate in the signal forwarding, the action space of each agent can be defined as follows.
\begin{equation}
    \mathcal{A}_t^k \triangleq [a_k(t)], 
\end{equation}
where $a_k(t) \in \{0, 1, 2, \dots, L\}$.	
\end{itemize}

Based on the global state $s_t \in \mathcal{S}_t \triangleq \mathcal{S}_t^1 \cup \mathcal{S}_t^2 \cup \dots \cup \mathcal{S}_t^K$, and the joint action $a_t \in A_t \triangleq \mathcal{A}_t^1 \cup \mathcal{A}_t^2 \cup \dots \cup \mathcal{A}_t^K$, the \textbf{reward} feedback to each agent can be defined as follows. 
\begin{equation} \label{reward function}
    r_t^k \triangleq r_t(s_{t+1}, a_t) \triangleq EE(a_t; s_t, p), 
\end{equation} 
which we refer to as `delayed reward'.
Because such reward is determined by the joint action in the current time slot (i.e., the vector of transmit power), and the global state in the next time slot (i.e., the unavailable global instantaneous CSI).

\subsection{DQN-Based MARL Solution} 
Based on (\ref{reward function}), 
we first define a new state-action value function with delayed reward, $\tilde{Q}$, to represent the expected return after performing action $a_t$ in state $s_t$.
\begin{equation} \begin{aligned} \label{delayed reward}
&\tilde{Q}(s_t, a_t; \theta) = \mathbb{E}\big[ R_t; s_t, a_t\big] \\
&=\int_{s_{t+1}} p(s_{t+1}|s_t,a_t) r_t(s_{t+1}, a_t) \mathrm{d} s_{t+1} + \gamma \int_{s_{t+1}} p(s_{t+1}|s_t,a_t) \\
&\times \int_{a_{t+1}} \pi(a_{t+1}|s_{t+1}) \tilde{Q}(s_{t+1}, a_{t+1}; \theta) \mathrm{d}a_{t+1} \mathrm{d}s_{t+1} \\
&= \mathbb{E}_{s_{t+1}}\big[r_t(s_{t+1}, a_t)\big] + \gamma\mathbb{E}_{s_{t+1}, a_{t+1}}\big[\tilde{Q}(s_{t+1}, a_{t+1}; \theta)\big], 
\end{aligned}\end{equation}
where $\theta$ represents DNN parameters. 
The above defined $\tilde{Q}$ is in a general form 
for long-period decisions in regular communication environments. 
It is also applicable for situations of one-step decisions in pure physical communication environments, where $p(s_{t+1}|s_t,a_t)=p(s_{t+1}|s_t)$, and $\tilde{Q}(s_{t+1}, a_{t+1}; \theta)$ on the right-hand side (RHS) of (\ref{delayed reward}) can be regarded as 0 because it is physically independent of $\tilde{Q}(s_t, a_t; \theta)$.

The underlying DQN algorithm uses a deterministic action policy, that is, the agent only chooses the best action which induces the maximal value of $\tilde{Q}$. 
Therefore, the Bellman iterative function for $\tilde{Q}$ can be further written as 
\begin{equation}\begin{aligned}
&\tilde{Q}(s_t, a_t; \theta)\\ 
&= \mathbb{E}_{s_{t+1}}\big[r_t(s_{t+1}, a_t)\big] + \gamma\mathbb{E}_{s_{t+1}}\big[ \max\limits_{a_{t+1}}\tilde{Q}(s_{t+1}, a_{t+1}; \theta)\big] \\
&=\mathbb{E}_{s_{t+1}}\big[r_t(s_{t+1}, a_t) + \gamma \max\limits_{a_{t+1}}\tilde{Q}(s_{t+1}, a_{t+1}; \theta)\big], 
\end{aligned}\end{equation}
and the agent's goal is to find the RL-based optimal action policy $\tilde{\pi}$, satisfying 
\begin{equation}\begin{aligned}
& \tilde{Q}(s_t, \tilde{a}_t; \theta, \tilde{a}_t= \tilde{\pi}(s_t)) 
= \max_{a_t} \tilde{Q}(s_t, a_t; \theta). \\ 
\end{aligned}\end{equation}
%
Accordingly, the temporal difference (TD) error is defined in (\ref{TD error}), and its squared form is usually taken as the training loss. 
\begin{equation}\label{TD error}
    TD(t) = r_t + \gamma \max\limits_{a_{t+1}} \tilde{Q}(s_{t+1}, a_{t+1}; \acute{\theta}) - \tilde{Q}(s_t, a_t; \theta), 
\end{equation}
where $\acute{\theta}$ is a group of network parameters in another separate target network.
It remains fixed when optimizing, and will be replaced by $\theta$ from the evaluate network after a period of training epochs. 

More specifically, for each agent $k$, sequential experience tuples $e_t^k=\{s_t^k, a_t^k, r_t^k, s_{t+1}^k\}$ can be obtained through the interaction with the external environment. 
Then, the tuples are stored in an experience replay buffer $\mathcal{B}^k=\{e_1^k, e_2^k, \dots, e_t^k\}$. 
To improve the accuracy when fitting $\tilde{Q}(s_t^k, a_t^k; \theta^k)$, a large amount of samples for evaluating $r_t^k + \gamma \max\nolimits_{a_{t+1}^k} \tilde{Q}(s_{t+1}^k, a_{t+1}^k; \acute{\theta}^k)$ is needed. 
Therefore, when training, a batch of experience will be sampled and used to minimize the following loss function: 
\begin{equation} \label{eq loss}
    L(\theta)=\mathbb{E}_{e_j^k\sim\mathcal{B}^k}\big[
        \big( TD^k(j) \big)^2
    \big].
\end{equation}
After carrying differentiate operations on a set of loss functions, the standard non-centered RMSProp optimization will be employed to update parameters in the evaluate network.

The pseudo-code of the proposed multi-agent DQN learning algorithm can be found in \textbf{Algorithm \ref{algo}}. 

\begin{algorithm}[htb]
    \caption{Delayed Reward-Based Multi-Agent DQN for Energy Efficiency Maximization in Relaying Networks}
    \label{algo}
    \begin{algorithmic}[1]
        \STATE Initialize the experience replay buffer of each agent $\mathcal{B}^k$.
        \STATE Initialize each agent's evaluate network $\theta^k$ and target network $\acute{\theta}^k=\theta^k$.
        \FOR{episode $t^{epi}=1,\dots,u_{max}^{epi}$}
            \STATE Initialize the communication environment.
            \FOR{time slot $t=1,2,\dots,t_{max}$}
                \FOR{each agent}
                    \STATE Observe the current partial CSI $s_t^k$. 
                    \STATE Determine a power level $a_t^k$ using the Epsilon-Greedy algorithm, then execute.
                    \STATE Observe the next partial state $s_{t+1}^k$.
                \ENDFOR
                \STATE Based on global state $s_{t+1}$ and joint action $a_t$, the environment feedback reward $r_t^k$ to each agent.
                \FOR{each agent}
                    \STATE Collect and save the current experience $e_t^k$ in $\mathcal{B}^k$.
                    \STATE Sample a batch of experience tuples from $\mathcal{B}^k$.
                    \STATE Minimize the loss in (\ref{eq loss}), and update the evaluate network $\theta^k$ using RMSProp optimization. 
                    \STATE Every $C$ steps reset $\acute{\theta}^k=\theta^k$.
                \ENDFOR
            \ENDFOR
        \ENDFOR
    \end{algorithmic}
\end{algorithm}

\subsection{Upper Bound Performance Analysis}
When only the outdated CSI is available, we have $s_{t+1} = [\textbf{H}_t, P_s]$, where $\textbf{H}$ represents the global CSI. 
On the other hand, according to Section~\ref{subsect game model}, game-theoretic approaches can obtain the optimal joint action $a_t^*$ in each time slot $t$ when $\textbf{H}_t$ is available. 
Thus, the achieved maximum utility by using game-theoretic approaches can be written as 
\begin{equation}\begin{aligned}
    EE^*(\textbf{P}^*) 
    &= r_t(s_{t+1}, a_t^*; a_t^* = \pi^*_\mathcal{G}(s_{t+1})) \\
    &\geq r_t(s_{t+1}, \tilde{a}_t; \tilde{a}_t=\tilde{\pi}(s_t)), 
\end{aligned}\end{equation}
which establishes a connection between the game-theoretic and MARL-based approaches.

Following the definition of $\tilde{Q}$ in (\ref{delayed reward}), we can construct the optimal Bellman iterative function in the form of the state-value function, to represent the discounted cumulative utilities achieved by game-theoretic approaches.  
\begin{equation}\begin{aligned}
    Q_\mathcal{G}(s_t, a_t^*) 
    & = r_t(s_{t+1}, a_t^*) + \gamma Q_\mathcal{G}(s_{t+1}, a_{t+1}^*) \\
    & \geq r_t(s_{t+1}, \tilde{a}_t) + \gamma \max\limits_{a_{t+1}}\tilde{Q}(s_{t+1}, a_{t+1}; \theta).
\end{aligned}\end{equation} 
Further, we have 
\begin{equation}
    \tilde{Q}(s_t, \tilde{a}_t; \theta) \leq \mathbb{E}_{s_{t+1}}\big[ Q_\mathcal{G}(s_t, a_t^*) \big], 
\end{equation}
which indicates that the utilities achieved by game-theoretic approaches provide the upper bounds for those achieved by MARL-based methods. 

Note that, because the above analysis is conducted directly on the general form of $\tilde{Q}$, 
the conclusion obviously holds for the situation of the investigated one-step decisions in pure physical communication environments. 

    \subsection{Complexity Analysis}
    Since the distributed agents act simultaneously, the time complexity of the MARL method can be estimated based on the DNN architecture of a single agent.
    The agent's network only has fully connected layers. 
    Suppose that the network has $M$ layers, and the $m$-th layer has $N_m$ neurons. 
    Then, similar to analysis in \cite{10374206}, the time complexity of deriving a group of actions is $\mathcal{O}(\sum_{m=1}^{M-1} N_m N_{m+1})$, where $N_1 = |\mathcal{S}_t^k|$ and $N_M = |\mathcal{A}_t^k|$. 
    This result also indicates the time complexity of the policy execution in the testing process. 
    In the training process, by further taking amounts of training episodes $u_{max}^{epi}$ and time slots in each episode $t_{max}$, we have the overall time complexity of \textbf{Algorithm \ref{algo}} as $\mathcal{O}(u_{max}^{epi} t_{max} \sum_{m=1}^{M-1} N_m N_{m+1})$.

\section{Numerical Evaluation}\label{sect Result}
In this section, we introduce simulation setups, and then evaluate the performance of the proposed methods. 
We set the channel state to change according to (\ref{channel state}) with parameter $\rho = 0.1$. 
The transmit power of the source is 1.0 W. 
The maximum transmit power of each relay is 1.0 W. 
The amount of relays is set to $K=5$. 
Similar to \cite{9568962}, we set the learning rate to 0.001 for each agent. 
The buffer size for storing experiences is 10000, and the mini-batch size for learning is 128. 
The interval that agents perform training is 1. 
For comparison, the following benchmarks are considered.
\begin{itemize}
\item \textbf{Optimal}: The optimal joint action is obtained by using game-theoretic approaches with global instantaneous CSI. 

\item \textbf{Single Agent DQN Scheme (SADQN)}: 
A central controller acts as the RL agent. 
It uses the DQN algorithm to learn the joint policy based on global outdated CSI, and schedules for all relays. 

\item \textbf{Global Outdated Information-Based Solution (GOIBS)}: 
The solution is derived through game-theoretic approaches under the global outdated CSI. 


\item \textbf{Partial Outdated Information-Based Solution (POIBS)}: 
Each relay $k$ only has the partially outdated CSI and tries to maximize the energy efficiency of the $s-k-d$ link. 
\end{itemize}

In the simulation, we set the total number of training episodes $u_{max}^{epi}=100$, and the
maximum number of time slots per episode $t_{max}=1000$. 
The hyper-parameter $\epsilon$ for the Epsilon-Greedy algorithm is fixed to 0.9 when training.

\begin{figure}[ht]
    \vspace{-0.4cm}
    \centering
    \includegraphics[scale=0.55]{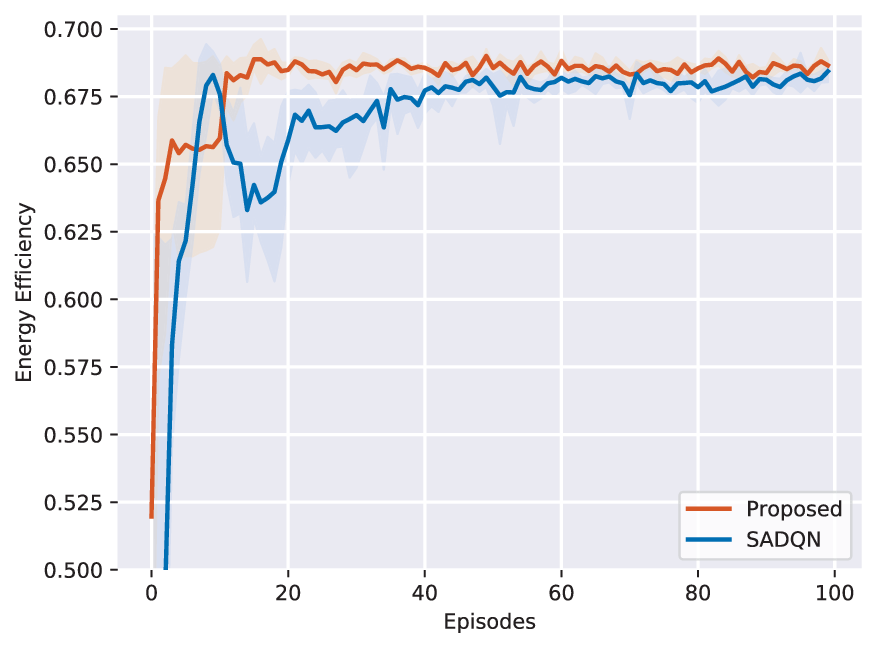}
    \caption{Performance of the distributed and centralized learning methods in the training stage.}
    \label{train}
    \vspace{-0.3cm}
\end{figure}

Consider that only the proposed MADQN and the SADQN method involve the learning process. 
Firstly, we set the total amount of power levels, $L+1$, to 4 as an example, and present the training performance of these two methods. 
The results are depicted in Fig.~\ref{train}, where dark lines represent the median of 5 successful trails, and shadow regions represent their range. 

We can find that the proposed method can converge faster and achieve better training stability than the SADQN method. 
Specifically, our method converges to a stable value after about the 40th episode, while the SADQN method converges after about the 60th episode. 
Moreover, the SADQN scheme undergoes stronger fluctuation during training. 
This is because the SADQN method uses a single agent to propose schemes for all relays. 
Since the DQN algorithm can only handle problems with one-dimensional actions, the combined actions from each relay will make the action space of SADQN very large, whose size is $(L+1)^K$. 
Consequently, it can be more difficult for the single agent to perform effective exploration and learn the optimal action policy efficiently, especially when facing the dimensional explosion of action space.
By contrast, the proposed method is a multi-agent scheme based on the DQN algorithm, where each agent only needs to deal with its original action space, whose size is $L+1$. 
Therefore, each agent can quickly explore and learn a good policy. 

On the other hand, the aforementioned problem also influences the final performance of the learned policy. 
It can be observed that the achieved energy efficiency of the proposed and SADQN methods are about 0.685 and 0.680 bit/J/Hz, respectively. 
Although the SADQN theoretically has the same upper bound of performance as ours, our method performs slightly better than the SADQN method after convergence.

\begin{figure}[ht]
    \vspace{-0.4cm}
    \centering
    \includegraphics[scale=0.55]{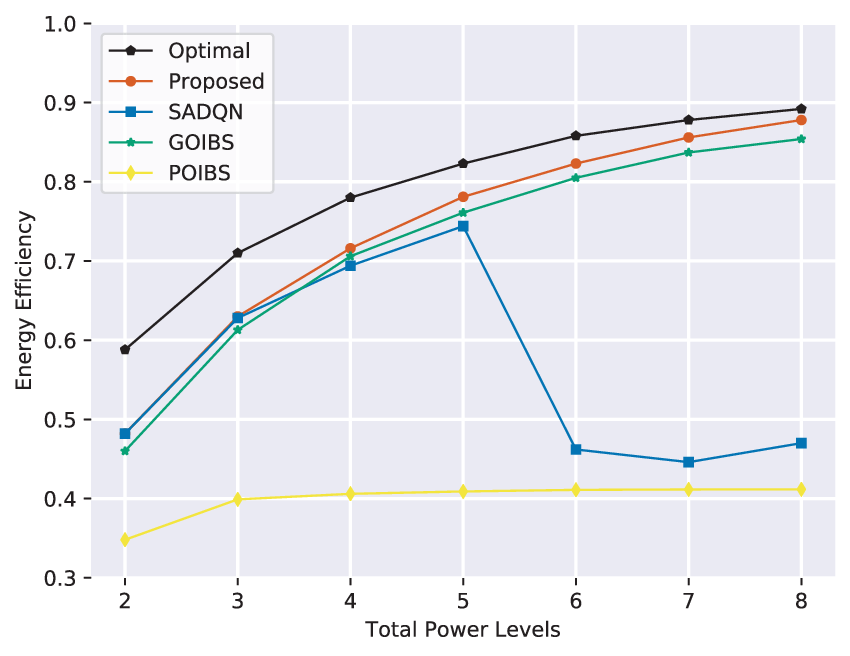}
    \caption{Performance of different methods under different total amounts of power levels at each relay.}
    \label{test power level}
    \vspace{-0.3cm}
\end{figure}

Further, we adjust the total amount of power levels at each agent ranging from 2 to 8, and compare our method with all the benchmarks. 
For the two learning approaches, i.e., the proposed and SADQN, $\epsilon$ is fixed to 1.0 when testing. 
That is, there is no random exploration in the testing stage, and RL agents directly determine actions according to the learned policies. 
The results are depicted in Fig.~\ref{test power level}.

It can be observed that all curves, except the SADQN's, increase as the total number of power levels increases. 
Among all the methods, the Optimal performs best, which also represents the upper-bound performance of the cooperative game. 
Take the situation of 5 total power levels as an example. 
With the global instantaneous CSI, the Optimal achieves the highest energy efficiency, which is about 0.823 bit/J/Hz. 
The achieved energy efficiency of our method is about 0.781 bit/J/Hz, only approximately 5.2\% away from the optimum. 
The performance gap is acceptable to some extent, considering that the instantaneous CSI may be difficult to obtain and the agents may not be able to achieve global information of the communication system. 

When the global information is available, but the instantaneous CSI is unavailable, the achieved energy efficiency of the GOIBS is about 0.761 bit/J/Hz, approximately 2.6 \% worse than ours. 
This is because the GOIBS directly uses outdated CSI to derive policies through game-theoretic approaches. 
In situations where the CSI in adjacent time slots varies greatly, the outdated CSI can be misleading and degrade final performance.
By contrast, our method can learn better policies, by implicitly modeling the transition dynamics of environment states and making distributed agents cooperate.

When only the partial and outdated CSI is available, the achieved energy efficiency of the POIBS is about 0.409 bit/J/Hz, which is much worse than ours.
This is because the non-learning POIBS method makes each relay only maximize the energy efficiency of its single forwarding link. 
The consequent global performance can hardly be optimal. 

In terms of the SADQN, the achieved energy efficiency under 5 total power levels is very close to ours. 
However, as analyzed for the training results in Fig.~\ref{train}, it suffers from the explosion of dimensions of combined actions. 
When the total amount of power levels keeps increasing, it shows a counter-intuitive performance decrease. 
This is because the size of SADQN's action space, i.e., $(L+1)^K$, can grow rapidly as $L$ increases. 
Despite that the optimal policy exists objectively, the SADQN can hardly find it. 
On the other hand, the explosion of dimension is avoided in the proposed multi-agent scheme, and better action policy can be learned in these tough cases.

\section{Conclusion}\label{sect Conclusion}
In this paper, we study energy efficiency maximization in non-ideal environments where only partially observable outdated CSI is available for each agent. 
To make RL methods suitable for the investigated environment, we define a delayed reward-based Q function for communication scenarios.
Then, we propose our MARL framework, and establish the connection between game-theoretic and MARL-based approaches by analyzing the upper bound performance of the proposed method. 
Simulation results reveal that our method outperforms existing learning and non-learning alternatives, with an acceptable deviation of about 5.2\% away from the optimum.

\bibliographystyle{IEEEtran}
\bibliography{paper_ref}

@ARTICLE{10857950,
	author={Zhang, Yuan and Wang, Rui and Liu, Erwu},
	journal={IEEE Transactions on Wireless Communications}, 
	title={Cooperative UAVs Placement Optimization for Best Multistatic Time-of-Arrival Localization in 5G Networks}, 
	year={2025},
	volume={24},
	number={4},
	pages={3561-3574},
}

@ARTICLE{10153599,
	author={Magalhães, Syllas R. C. and Bayhan, Suzan and Heijenk, Geert},
	journal={IEEE Transactions on Green Communications and Networking}, 
	title={Impact of Power Consumption Models on the Energy Efficiency of Downlink NOMA Systems}, 
	year={2023}, month={Jun.},
	volume={7},
	number={4},
	pages={1739-1753},
}

@ARTICLE{9675818,
	author={Guo, Yuting and Liu, Xin and Liu, Xueying and Durrani, Tariq S.},
	journal={IEEE Networking Letters}, 
	title={Energy-Efficient Resource Allocation for Simultaneous Wireless Information and Power Transfer in GFDM Cooperative Communications}, 
	year={2022}, month={Mar.},
	volume={4},
	number={1},
	pages={1-5},
}

@ARTICLE{9645220,
	author={Lv, Suyu and Xu, Xiaodong and Han, Shujun and Tao, Xiaofeng and Zhang, Ping},
	journal={IEEE Transactions on Vehicular Technology}, 
	title={Energy-Efficient Secure Short-Packet Transmission in NOMA-Assisted mMTC Networks With Relaying}, 
	year={2022}, month={Feb.},
	volume={71},
	number={2},
	pages={1699-1712}
}

@INPROCEEDINGS{9838809,
	author={Galkin, Boris and Omoniwa, Babatunji and Dusparic, Ivana},
	booktitle={ICC 2022 - IEEE International Conference on Communications}, 
	title={Multi-Agent Deep Reinforcement Learning For Optimising Energy Efficiency of Fixed-Wing UAV Cellular Access Points}, 
	year={2022}, month={Aug.},
	address={Seoul, Korea},
	pages={1-6},
}

@ARTICLE{10374206,
  author={Geng, Yuanzhe and Liu, Erwu and Wang, Rui and Sun, Pengcheng and Lu, Binyu and Wang, Jie},
  journal={IEEE Internet of Things Journal}, 
  title={Reinforcement-Learning-Based Policy Design for Outage Minimization in DF Relaying Networks}, 
  year={2024},
  volume={11},
  number={9},
  pages={15359-15374},
}

@ARTICLE{5710995,
	author={Suraweera, Himal A. and Tsiftsis, Theodoros A. and Karagiannidis, George K. and Nallanathan, Arumugam},
	journal={IEEE Transactions on Vehicular Technology},
	title={Effect of Feedback Delay on Amplify-and-Forward Relay Networks With Beamforming},
	year={2011},
	month={Mar.},
	volume={60},
	number={3},
	pages={1265-1271},
}

@ARTICLE{9568962,
	author={Geng, Yuanzhe and Liu, Erwu and Wang, Rui and Liu, Yiming},
	journal={IEEE Transactions on Communications},
	title={Hierarchical Reinforcement Learning for Relay Selection and Power Optimization in Two-Hop Cooperative Relay Network},
	year={2022},
	month={Jan.},
	volume={70},  number={1},  pages={171-184},
}

@ARTICLE{6954557,
	author={Shams, Farshad and Bacci, Giacomo and Luise, Marco},
	journal={IEEE Transactions on Wireless Communications}, 
	title={Energy-Efficient Power Control for Multiple-Relay Cooperative Networks Using  $Q$-Learning}, 
	year={2015}, month={Mar.},
	volume={14},
	number={3},
	pages={1567-1580},
}

@book{Osborne1997game,
	title={A Course in Game Theory},
	author={M. J. Osborne and A. Rubinstein},
	year={1997},
	publisher={Cambridge, MA, USA: MIT Press}, 
}

@ARTICLE{Nash1950,
	title={Equilibrium points in n-person games},
	author={J. F. Nash},
	year={1950}, month={Jan.}, 
	publisher={Proc. Nat. Academy Sci. United States America}, 
	vol={36},
	no={1},
	pp={48–49}
}

\end{document}